\documentclass[preprint,amsmath,amssymb]{revtex4}

\usepackage{epsfig}

\begin{document}

\title{Electro-viscous effects in capillary filling of nanochannels}

\author{Niels Asger Mortensen\footnote{asger@mailaps.org} and Anders Kristensen}%

\affiliation{Department of Micro and Nanotechnology, NanoDTU,
Technical University of Denmark, bldg.~345 east, DK-2800 Kongens
Lyngby, Denmark}

\date{\today}

\begin{abstract}
We theoretically examine the widespread hypothesis of an
electro-viscous origin of the increase in apparent viscosity
observed in recent experiments on capillary filling of nanochannels.
Including Debye-layer corrections to the hydraulic resistance we
find that the apparent viscosity reaches a maximum in the mesoscopic
regime where the channel height (or more generally the
hydraulic radius) is comparable to the screening length. However,
for realistic estimates of central parameters, we find that the
electro-viscous contribution to the apparent viscosity is at most a
$1\%$ effect.
\end{abstract}
\maketitle

Passive nanofluidic devices have recently been explored in a number
of contexts aiming at both fundamental studies such as DNA
stretching~\cite{Tegenfeldt:2004,Reisner:2005,Fu:2006,Reisner:2007,Liang:2007}
as well as for applications in optofluidic dye
lasers~\cite{Gersborg-Hansen:2006}. During the past couple of years
this has also stimulated basic experimental investigations of fluid
dynamics particularly addressing possible new phenomena associated
with the cross-over from a macroscopic (i.e. micrometer scale) to
the nano regime. The list relevant length scales and phenomena
associated with this cross over includes the Debye screening length
$\lambda_D$ (typically of the order 10 nm), the zeta surface
potential $\zeta$ (typically sub-$k_BT$ and of the order meV), and
the issue of slip versus no-slip on the inter-molecular
sub-nanometer length scale $\delta$~\cite{Bruus:2007}.

Of particular importance is to achieve a more complete
understanding of fluid dynamics in the \emph{mesoscopic} regime,
where relevant length scales of the fluid, e.g. the Debye
screening length, is comparable to device dimensions, for example
the channel height $h$, while still exceeding the molecular length
scale, i.e. $\delta\ll \lambda_D\lesssim h$. This mesoscopic
regime is typically entered when capillaries with cross sectional
dimensions of approximate 100~nm and below, is infilled with
buffer solutions with an ionic strength relevant for carrying
typical biological samples, e.g. DNA. A typical salinity in the mM
range results in $\lambda_D\sim 10$\,nm~\cite{Bruus:2007}.

\begin{figure}
\centerline{
\epsfig{file=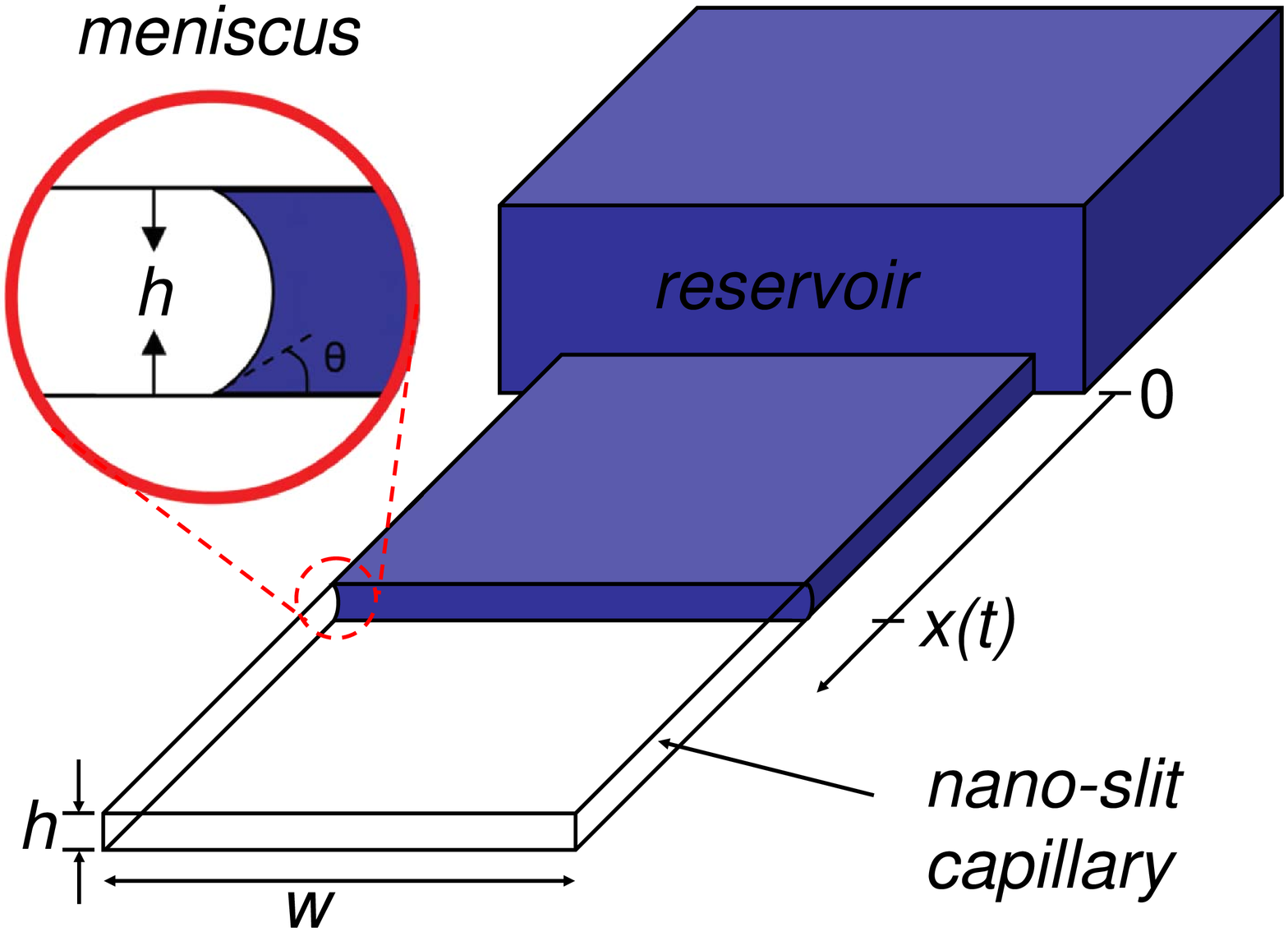,width=\columnwidth,clip}
} \caption{(Color online) Schematic
illustration of a nanoslit-capillary filling experiment. A
nanoslit capillary, of height $h \leq 300$~nm and width $w$
several $\mathrm{\mu}$m is filled with liquid from an attached
reservoir by capillary action. The length of the fluid plug - the
position of the liquid meniscus - $x(t)$ is recorded as function
of time $t$.} \label{fig1}
\end{figure}

This has motivated several basic nano-scale experimental
investigations on the classical problem of capillary filling, see
Fig.~\ref{fig1}, thus testing the classical Washburn
formula~\cite{Wasburn:1921}
\begin{equation}\label{eq:Washburn}
x^2(t)=\frac{\gamma h\cos\theta}{3\eta}t
\end{equation}
which is known to provide a well-established description of the
filling dynamics in capillaries with cross-sectional dimensions of
mm size and down below 1\,$\rm \mu m$. Here, $x$ is the length of
advancement along the channel of the liquid front, $\gamma$ is the
water-air surface tension, $\theta$ is the contact angle, and
$\eta$ is the viscosity. The Washburn formula is derived within
the framework of the Stokes equation for a charge-neutral liquid
where no surface-charging and electro-osmotic phenomena occur.
Eq.~(\ref{eq:Washburn}) implicitly relies on the continuum
hypothesis which limits its validity for fluid dynamics at the
inter-molecular length scale while it is still expected to work in
the mesoscopic regime where physical mean-fields remain
well-defined and continuum formalism applies.

Experimental observations support an unexpected increase in the
filling time, beyond the predictions of Eq.~(\ref{eq:Washburn}),
in shallow nanoslit devices for slit heights of approximate 100~nm
and below~\cite{Tas:2004, Thamdrup:2007}. In all reported
experiments, a systematic deviation from the behavior predicted by
the Washburn formula is observed: Although the observed dynamics
maintain the proportionality between $x^2$ and $t$, the constant
of proportionality $a$ does not scale linearly with the channel
height $h$ for slit channel heights of approximate 100~nm and
below.

In absence of a detailed theoretical model, different possible
interpretations have been presented for the unexpected height
dependence of $a(h)$. Tas \emph{et al.} speculate that the
phenomena is of an electro-viscous origin~\cite{Tas:2004} while
Thamdrup \emph{et al.} show that in their experiments the
phenomena correlates with the formation of air
bubbles~\cite{Thamdrup:2007}. Both interpretations implicitly
assume that changes in hydraulic resistance $R_{\rm hyd}$ play a
central role. For this reason, the changes in observed filling
time are naturally presented in terms of an increase in apparent
viscosity $\eta$, entering the Washburn formula.

\begin{figure}
\centerline{
\epsfig{file=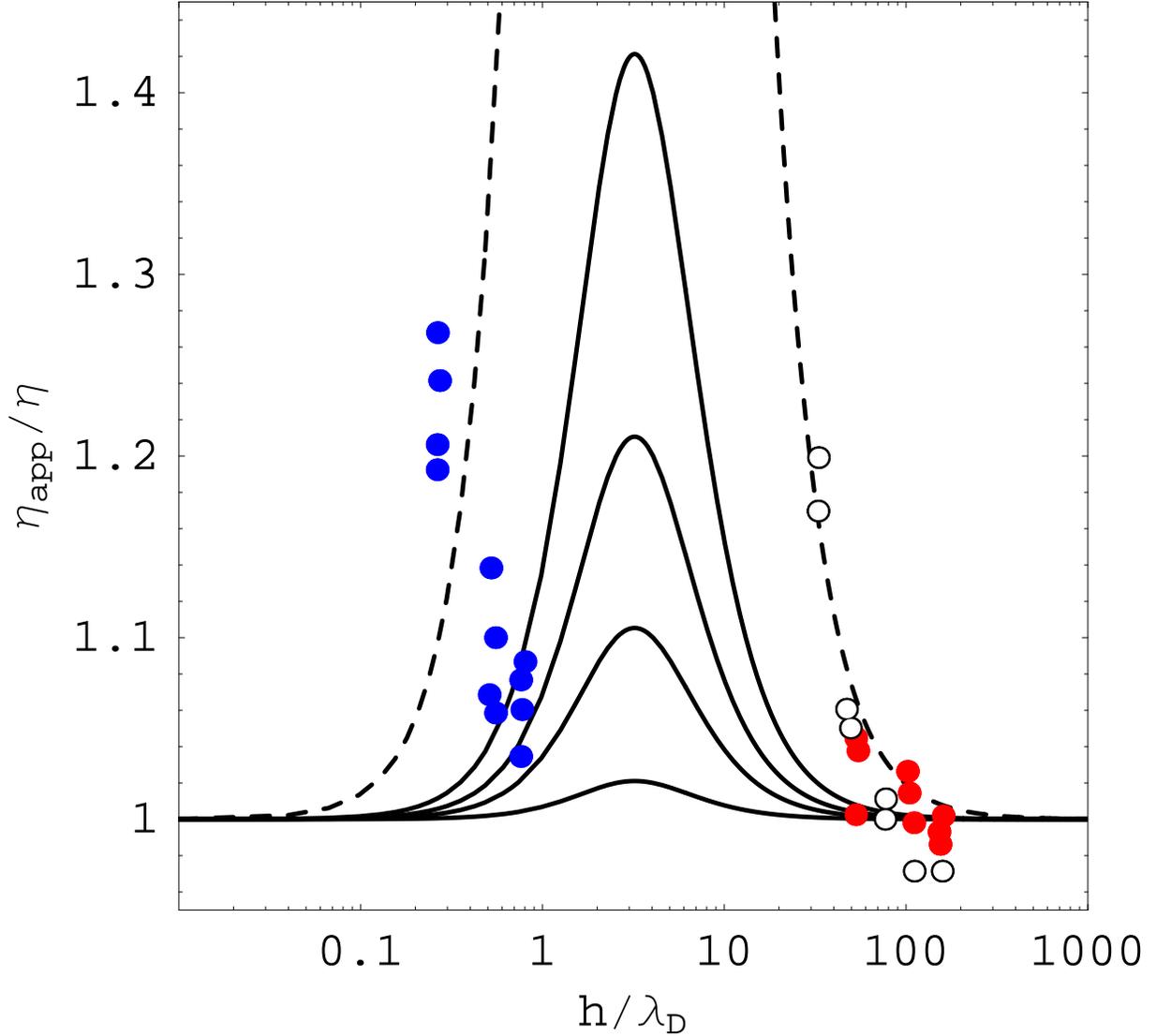,width=\columnwidth,clip}
} \caption{(Color online) Plot of the
apparent viscosity $\eta_{\rm app}/\eta$ versus normalized channel
height $h/\lambda_D$ for increasing values of the the ratio
between the electro-osmotic mobility and the viscous mobility
$\chi= 0.1, 0.5, 1$, and $2$ (solid lines from below). The solid
data points show experimental results reproduced from
Ref.~\cite{Tas:2004} for demi water with $\lambda_D\simeq 200$\,nm
(left group of data) and 0.1M NaCl solution with $\lambda_D\simeq
1$\,nm (right group of data). The open data points show
corresponding results for 0.1M NaCl reproduced from
Ref.~\cite{Persson:2007} with the dashed line being a fit to the
data with $\chi\sim 16$ used as fitting parameter.} \label{fig2}
\end{figure}

In this Letter we work out the electro-viscous correction to the
apparent viscosity $\eta_{\rm app}$ and discuss the result and
consequences in the context of the experiments in
Refs.~\cite{Tas:2004,Thamdrup:2007}. Our theoretical account of the
electro-viscous correction includes the Debye-layer correction to
the hydraulic resistance~\cite{Mortensen:2006,Mortensen:2007}. When
decreasing the height $h$ the apparent viscosity increases in
qualitative agreement with the experiments, reaching a maximum when
the height of the channel $h$ is comparable to the Debye screening
length $\lambda_D$, and then drops off again when further reducing
the height beyond the Debye screening length, see Fig.~\ref{fig2}.
This picture agrees with earlier theoretical work on retarding
electrokinetic counter flow~\cite{Burgreen:1964}. However, when
scaling the experimental results with the Debye screening length the
quantitative agreement with experiments is poor and the prediction
of a maximum is not consistent with
experiments~\cite{Tas:2004,Thamdrup:2007}. This suggest that there
are other phenomena contributing to the apparent viscosity, thus
supporting the explanation by the formation of air
bubbles~\cite{Thamdrup:2007} over the electro-viscous
hypothesis~\cite{Tas:2004}.

The basic physics behind the electro-viscous hypothesis has
already been discussed by Tas \emph{et al.}~\cite{Tas:2004}, but
here we repeat the arguments for completeness. When water and
other electrolytes flow in a channel, chemically induced
charge-transfer takes place at the channel walls rendering a
charged Debye layer of width $\lambda_D$ in the liquid near the
wall, and a compensating oppositely charged wall. This electric
phenomena combines with the purely viscous flow and results in
electro-viscous effects. In the spirit of the derivation of the
Washburn formula~\cite{Wasburn:1921}, the apparent viscosity of
course stems from an apparent increase in hydraulic resistance
$R_{\rm hyd}$ due to an electro-osmotic counter flow. Since the
hydraulic resistance scales linearly with the viscosity we have
the following for the Debye-layer correction to the apparent
viscosity
\begin{equation}\label{eq:etaapp}
\frac{\eta_{\rm app}}{\eta}=\frac{R_{\rm hyd}(\zeta)}{R_{\rm
hyd}(\zeta=0)}
\end{equation}
where $\zeta$ is the zeta-potential at the surface of the channel.

The corrections from a finite zeta potential follows from our
recent general linear-response theory of combined mass and charge
transport in long straight
channels~\cite{Mortensen:2006,Mortensen:2007} with the electrolyte
being subject to the application of an external pressure drop
$\Delta p$ and voltage drop $\Delta V$ along the channel. These
potential drops results in a liquid flow rate $Q$ and an electric
current $I$ given by the $2\times2$ conductance matrix $G$ as
\begin{equation} \label{eq:Gmatrix}
\begin{pmatrix}Q\\I\end{pmatrix} =
\begin{pmatrix} G_ {11} & G_ {12} \\
G_ {21} & G_ {22}\end{pmatrix}
\begin{pmatrix} \Delta p\\ \Delta V\end{pmatrix},
\end{equation}
with the matrix being subject to an Onsager relation,
$G_{12}=G_{21}$. In the absence of a zeta potential we of course
have $G_{12}(\zeta=0)=G_{21}(\zeta=0)=0$ so that
Eq.~(\ref{eq:etaapp}) becomes
\begin{equation}\label{eq:etaapp2}
\frac{\eta_{\rm app}}{\eta}
=\left[1-\frac{G_{12}G_{21}}{G_{11}G_{22}}\right]^{-1},
\end{equation}
where we have used that $R_{\rm hyd}(\zeta)=R_{11}(\zeta)$ with
the resistance matrix $R=G^{-1}$ given by the inverted conductance
matrix. In applying the result in Eq.~(\ref{eq:Gmatrix}) to the
present problem of capillary filling we have implicitly utilized
that the time scale for charge-transfer and charging of the Debye
layer is several orders of magnitude smaller than the time-scale
for moving the meniscus forward the distance of one Debye length.
This difference in time scales secures that at any instant the
charge distribution has reached its equilibrium, and consequently
the electric current is zero, $I=0$.

We now proceed by calculating the conductance matrix elements by
analytical means. For $h\ll w$ the problem becomes quasi
one-dimensional and within the Debye--H\"uckel limit it can be
solved exactly. In particular we have
\begin{subequations}
\begin{align}
G_{11}&=\frac{h^3w}{12L\eta },\\
G_{12}&=G_{21}= -\frac{hw}{L}\frac{\epsilon \zeta}{\eta} \left[1-
\frac{2\lambda_D}{h} \tanh\left(\frac{h}{2\lambda_D}\right)\right],\\
G_{22}&= \frac{hw}{L}\sigma- \frac{\epsilon^2\zeta^2 hw }{\eta L
\lambda_D^2}\frac{1-\frac{\lambda_D}{h}\sinh\left(\frac{h}{\lambda_D}\right)}{1+\cosh\left(\frac{h}{\lambda_D}\right)},
\end{align}
\end{subequations}
where $L$ is the length of the channel, $\sigma=\epsilon
D/\lambda_D^2$ is the conductivity of the liquid, $D$ is the
diffusion constant of the ions, and $\epsilon$ the dielectric
function of the liquid. Long, but tedious manipulations then leads
to
\begin{equation}\label{eq:final}
\frac{\eta_{\rm app}}{\eta}= 1 + 12
\left(\frac{\lambda_D}{h}\right)^4\left(\frac{\frac{h}{\lambda_D}-2+\left(\frac{h}{\lambda_D}+2\right)
e^{-h/\lambda_D}}{1+e^{-h/\lambda_D}}\right)^2\chi
\end{equation}
which is correct to lowest order in the the dimensionless
parameter
\begin{equation} \label{eq:chidef}
\chi\equiv
\frac{\epsilon^2\zeta^2}{\eta\sigma\lambda_D^2}=\frac{\epsilon\zeta^2
}{\eta D } = \frac{\eta \mu_\mathrm{eo}^2}{\epsilon D}
\end{equation}
expressing the square of the ratio between the electro-osmotic
mobility and the viscous mobility. The latter form of $\chi$ is
perhaps most appropriate from an experimental point of view, as
the electro-osmotic mobility $\mu_\mathrm{eo}=\epsilon\zeta/\eta$
can be determined by measurements easier than the zeta potential
$\zeta$ itself.

In Fig.~\ref{fig2} we show the apparent viscosity for increasing
values of $\chi$. Most importantly it should be emphasized that
the electro-viscous correction is maximal for $h\simeq
3.212\lambda_D$ which is a value that can be found numerically.
This is the mesoscopic regime where the Debye screening length
$\lambda_D$ becomes comparable to the channel height $h$, while
still exceeding the inter-molecular length scale $\delta$.

The plot also includes experimental data reproduced from
Refs.~\cite{Tas:2004,Persson:2007} for both a 0.1M NaCl solution
with $\lambda_D\sim 1$\,nm~\cite{Tas:2004,Persson:2007} as well as
for demi water with $\lambda_D\sim 200$\,nm~\cite{Tas:2004}. While
the experiments with a NaCl solution belongs to the sup-mesoscopic
regime with $h\gg\lambda_D$ the demi-water experiments are in the
opposite sub-mesoscopic limit with $h<\lambda_D$. As seen, the
demi-water results are not consistent with the hypothesis of a
correction of an electro-vicious origin as derived above. We note
that this conclusion is quite solid, since the demi-water data
would belong to the sub-mesoscopic regime (to the left of the
peak) even if we for some reason would allow for an one-order of
magnitude reduction in the estimate of the screening length, i.e.
$\lambda_D\sim 200\,{\rm nm}\longrightarrow 20\,{\rm nm}$. The
dashed line shows a fit of Eq.~(\ref{eq:final}) to the NaCl data
with $\chi\sim 16$ used as fitting parameter. However, we
emphasize that this high value of $\chi$ is completely unrealistic
and in fact for a zeta potential in the meV range we find that
$\chi\lesssim 10^{-3}$ thus leaving the estimated relative
electro-viscous correction a sub-1\% effect at most.

Nanofluidic capillaries have recently been used in
optofluidics~\cite{Gersborg-Hansen:2006} and in applications for
stretching of DNA
molecules~\cite{Tegenfeldt:2004,Reisner:2005,Fu:2006,Reisner:2007,Liang:2007}.
The majority of these experiments employ nanochannels with roughly a
square cross section. The above quasi 1D results may be generalized
to more arbitrary 2D channel cross sections with an aspect ratio of
order unity. Following Ref.~\onlinecite{Mortensen:2007} we
straightforwardly arrive at the approximate result
\begin{equation}\label{eq:final2}
\frac{\eta_{\rm app}}{\eta}\simeq 1 +
\frac{8I_2^2\left(\frac{R}{\lambda_D}\right)}{\left(\frac{R}{\lambda_D}\right)^2I_0^2\left(\frac{R}{\lambda_D}\right)}\chi
\end{equation}
where $I_n$ is the modified Bessel function of the first kind of
order $n$ and $R=2A/P$ is the hydraulic radius with $A$ being the
cross-sectional area and $P$ being the length of the perimeter. For
a circular cross section with radius $a$ we have $R=a$ and
Eq.~(\ref{eq:final2}) is exact in the limit of a small $\chi$.
Likewise, for a square channel with side length $a$ we get $R=a/2$.
Eq.~(\ref{eq:final2}) gives rise to the same qualitative dependence
on $R/\lambda_D$ as in Fig.~\ref{fig2} and our above estimates and
conclusions remain unchanged for nano channels with a true 2D
electro-hydrodynamic flow profile.

While we are confident that the electro-viscous effect would reveal
itself in a clean experiment (i.e. in the absence of other competing
effects) with a true nano-scale slit ($h\sim 3\lambda_D$) the
strength of the effect is far too weak to account for the
experimental observations. This strongly supports the existence of
other dominating effects, as also observed in recent studies where
the apparent viscosity was found to correlate with the presence of
air bubbles which tend to increase the hydraulic
resistance~\cite{Thamdrup:2007}.

In conclusion, our calculation of the Debye-layer correction shows
that for realistic estimates of central parameters, the
electro-viscous contribution to the apparent viscosity is below a
$1\%$ level.

 \emph{Acknowledgements}. We thank H. Bruus
for stimulating discussions during the initial phase of the
project.


\newpage

\end{document}